\documentclass[12pt,letterpaper,twoside]{article}

\usepackage{graphicx}

\begin{document}

\newcommand {\be} {\begin{equation}}
\newcommand {\ee} {\end{equation}}
\newcommand {\bea} {\begin{eqnarray}}
\newcommand {\eea} {\end{eqnarray}}
\newcommand {\eq} [1] {Eq.\ (\ref{#1})}
\newcommand {\eqs} [2] {Eqs.\ (\ref{#1}) and (\ref{#2})}
\newcommand {\Eq} [1] {Equation (\ref{#1})}
\newcommand {\Eqs} [2] {Equations (\ref{#1}) and (\ref{#2})}
\newcommand {\fig} [1] {Fig.\ \ref{#1}}
\newcommand {\Fig} [1] {Figure \ref{#1}}
\newcommand {\figs} [2] {Figs.\ \ref{#1} and \ref{#2}}
\newcommand {\Figs} [2] {Figures \ref{#1} and \ref{#2}}

\def \eps {\epsilon}
\def \veps {\varepsilon}
\def \pl {\partial}
\def \hf {{1 \over 2}}
\def \mf {\mathbf}

\def\lsim{\mbox{\raisebox{-.6ex}{~$\stackrel{<}{\sim}$~}}}
\def\gsim{\mbox{\raisebox{-.6ex}{~$\stackrel{>}{\sim}$~}}}


\title{\bf Testing for Features in the Primordial Power Spectrum}

\author{
\normalsize Loison~Hoi and James~M.~Cline\\
\small \it Department of Physics, McGill University\\
\small \it 3600 Rue University, Montr\'eal, Qu\'ebec, Canada H3A 2T8\\
\small E-mail: hoiloison@physics.mcgill.ca, jcline@physics.mcgill.ca}

\date{\small December 22, 2009}

\maketitle

\begin{abstract}
Well-known causality arguments show that events occurring  during or
at the end of inflation, associated with reheating or preheating,
could contribute a blue component to the spectrum of primordial
curvature perturbations, with the dependence $k^3$.    We explore the
possibility that they could be observably large in CMB, LSS,
and Lyman-$\alpha$ data. We find that a $k^3$ component with a cutoff
at some maximum $k$ can modestly improve the fits
($\Delta\chi^2=2.0, 5.4$) of the low multipoles ($\ell \sim 10 - 50$)
or the second peak ($\ell \sim 540$) of the CMB angular spectrum when
the three-year WMAP data are used. Moreover, the results from WMAP
are consistent with the CBI, ACBAR, 2dFGRS, and SDSS data when they
are included in the analysis. Including the SDSS galaxy clustering
power spectrum, we find weak positive evidence for the $k^3$
component at the level of  $\Delta{\chi^2}^\prime = 2.4$, with the
caveat that the nonlinear evolution of the power spectrum may not be
properly treated in the presence of the  $k^3$ distortion.  To
investigate the high-$k$ regime, we use the Lyman-$\alpha$ forest
data (LUQAS, Croft {\it et al}., and SDSS Lyman-$\alpha$);  here we
find evidence at the level $\Delta{\chi^2}^\prime = 3.8$.
Considering that there are two additional free parameters in the
model, the above results do not give a strong evidence for features;
however, they show that surprisingly large bumps are not ruled out.
We give
constraints on the ratio between the $k^3$ component and the nearly
scale-invariant component, $r_3 < 1.5$, over the range of
wave numbers $2.3\times10^{-3}\ {\rm Mpc}^{-1} < k < 8.2\ {\rm
Mpc}^{-1}$. We also discuss theoretical models which could lead to
the $k^3$ effect, including ordinary hybrid inflation and double
D-term inflation models. We show that the well-motivated $k^3$ component is also a good representative of the generic spikelike feature in the primordial perturbation power spectrum.

\end{abstract}


\section{Introduction}

Inflation has become a cornerstone of modern cosmology; it not only
solves critical cosmological problems, but also provides
possibilities of exploring the infant universe
\cite{Guth, inflation1, inflation2}. A crucial aspect of inflationary
theory is the primordial perturbation power spectrum, which connects
the quantum fluctuations in the early universe to the formation of
structure at later times. Inflation generically predicts a
nearly scale-invariant primordial power spectrum, which agrees well
with cosmic microwave background (CMB), large-scale structure (LSS),
and Lyman-$\alpha$ forest observations, such as the Wilkinson
Microwave Anisotropy Probe (WMAP) \cite{WMAP1-1, WMAP3-1}, Cosmic
Background Imager (CBI) \cite{CBI}, Arcminute Cosmology Bolometer
Array Receiver (ACBAR) \cite{ACBAR}, Two-degree-Field Galaxy Redshift
Survey (2dFGRS) \cite{2dF, 2dF2005}, Sloan Digital Sky Survey (SDSS)
\cite{SDSS, Tegmark:2006az}, and Lyman-$\alpha$ forest (Viel {\it et al}.\
\cite{Viel, Kim, Croft} and SDSS Lyman-$\alpha$ \cite{sdsslya1,
sdsslya2}).

While producing a nearly scale-invariant spectral index has become a
major criterion of model selection, a deviation from pure
scale invariance is consistent with the current data, and in fact
fits better than $n_s=1$. Further elaborations have been 
investigated; a well-known example, suggested by the first-year WMAP
data (WMAP1) \cite{WMAP1-2}, is the running spectral index model,
which has a large and negative running of the spectral index and a
large tensor-to-scalar ratio (although the evidence for this is
weakened when the three-year WMAP data [WMAP3] is used
\cite{WMAP3-1}). Detailed investigation revealed that a partially
running spectral index model, which has a constant spectral index on
large and small scales but a running spectral index on the relevant
scales (about 3.4 e-foldings), provides as good a fit to WMAP1 as the
full running spectral index model \cite{running}.  Because of the
limited range of scales, this could be described as adding a
localized {\it feature} to the primordial spectrum, a topic which has
also received considerable attention 
\cite{Starobinsky:1992ts, Adams:2001vc, Gong:2005jr, Kawasaki:2006zv}, and which is 
the theme of the present work.

Other hints of peculiarities in the CMB data have prompted the
investigation of spectral features going beyond a simple power law.
For example, the reconstruction of the primordial power spectrum
directly from the CMB angular spectrum data indicates the prominent
feature of an infrared cutoff on the horizon scale \cite{pkrecon,
pkrecon2}. There are also a few outlying multipoles at smaller
scales, $\ell \sim 10 - 50$ and $\ell \sim 540$, which have inspired
modifications to the spectrum, such as would be provided by 
introducing a sharp step in the inflaton potential
\cite{WMAP1-2, Starobinsky:1992ts, Adams:2001vc, Gong:2005jr}.  This introduces oscillations into the
spectrum  which allow for better fits to the data
\cite{future, Hamann:2007pa}.
More recent analyses of this subject can be found in  
Refs.\ \cite{TocchiniValentini:2004ht, Hunt:2004vt, Hunt:2007dn, Shafieloo:2007tk, Joy:2007na, Verde:2008zza, Pahud:2008ae, Joy:2008qd, Shafieloo:2009cq, Paykari:2009ac, Nicholson:2009pi, Mortonson:2009qv}.
Such improvements of course come at
the price of adding more parameters to the model; questions of
significance can be handled through  quantitative model selection
criteria \cite{Liddle, Pahud:2006kv, Liddle:2006tc, Liddle:2007fy}.

Because the current data do not justify adding a large number of
parameters to the description of the power spectrum, it is important
to consider models with a small number of extra parameters, and 
preferably with a strong theoretical motivation.   In the present
work, we will investigate a new kind of spectral feature in this
category, namely the addition of a component which scales like $k^3$,
in contrast to the nearly scale-invariant spectrum  $\sim k^{0}$. 
As we will describe in Section \ref{motivation}, this behavior was a
generic prediction \cite{Abbott-Traschen} based on the requirement of
causality, prior to the invention of inflation.  Our observation is
that extra contributions to the spectrum arising after inflation, for
example during reheating, could be expected to scale like $k^3$ at
low $k$.  In fact, it was recently shown that this effect can arise
in hybrid inflation models \cite{k3, k3_2} and double D-term inflation models \cite{julien1}.

Of course, other qualitatively similar distortions could arise in
particular inflationary models, having $k^n$ spectra with $n\neq 3$. 
In this paper we focus on the $n=3$ case because of the theoretical
motivations mentioned above, and upon which we will elaborate further
below.   However, for much of our analysis of the data, we would not
expect greatly different results for nearby values like $n=2$ or
$n=4$, for the reason that the large-$k$ behavior must be cutoff at
some scale $k_{\rm c}$; then the feature of interest is a localized spike
(since the data do not allow for larger features) which would exceed
the underlying scale-invariant contribution to ${\cal P}_{\cal R}(k)$ only over a
rather limited range of $k$.  In that sense, the $k^3$ model with a
cutoff can be considered as a rather generic model for spikelike
features in the primordial power spectrum.
Nevertheless, we will compare the $k^3$ spike to $k^n$ spikes with 
$n=1, 2, 4, 5$, which will clarify this point. 
Spikelike features in the primordial power spectrum have
been previously investigated in the literature. 
For example, Ref.\ \cite{Kawasaki:2006zv} finds a strong spike 
in the smooth hybrid inflation model; its location, however, 
is beyond the current CMB-LSS scales and hence is not observable.
In any case, we take the point of view that the $k^3$
model (with a cutoff) is sufficiently well-motivated to provide an interesting test
for a feature of this kind, and we have done a state-of-the-art comparison of this model of spectral features with the CMB, LSS, and Lyman-$\alpha$ data.

To describe the $k^3$ effect, we modify the primordial curvature
power spectrum, ${\cal P_R}(k) \equiv (L/2\pi)^34\pi k^3 \langle\left|{\cal R}_{\mf k}\right|^2\rangle$ (see Appendix \ref{appendixa}), 
to the form
\bea 
\label{P_Rk3} 
   {\cal P_R}(k) &=& {\cal P}_\phi(k) + {\cal
   P}_3(k)\nonumber\\ &=& {\cal P}_\phi(k_0) \left( {k \over k_0}
   \right)^{n_s-1} + {\cal P}_3(k_0) \left( {k \over k_0}
   \right)^3\nonumber\\ &=& {\cal P}_\phi(k_0) \left[ \left( {k \over k_0}
   \right)^{n_s-1} + r_3(k_0) \left( {k \over k_0} \right)^3
   \right], 
\eea 
where $k_0$ is a pivot point which we take to be $k_0=0.002$
Mpc$^{-1}$ throughout
(to be consistent with the WMAP collaboration).
The amplitude ratio of the $k^3$ component to the nearly scale-invariant component is 
\be r_3(k)
   \equiv {{\cal P}_3(k) \over {\cal P}_\phi(k)}. 
\ee 
We refer to the model whose power spectrum is described by \eq{P_Rk3}
as the $k^3$ model.  It has just one free parameter relative to the
power-law $\Lambda$CDM model: the ratio $r_3$.  However, in any realistic model,
the $k^3$ behavior must be cut off at some maximum scale $k_{\rm c}$; for
example, in tachyonic preheating, where the tachyon curvature $m_\sigma^2$ is
negative, only the perturbations satisfying  $k^2< a^2 |m_\sigma^2|$ are
amplified.  For simplicity we introduce a sharp cutoff: if $k>k_{\rm
c}$, then the $k^3$ term in \eq{P_Rk3} is set to 0.  We refer to this
as the $k^3_{\rm c}$ model.  It has one more parameter, $k_{\rm c}$. 

In this paper, we examine the extent to which the $k^3_{\rm c}$ model
is consistent with the data. We begin in Section \ref{motivation}
by reviewing the theoretical motivations for this effect.
Section \ref{data} continues with
a brief description of the CMB, LSS, and Lyman-$\alpha$ data used in
the analysis.  We investigate the evidence from WMAP in Section
\ref{fittingwmap}, and explore the $k^3$ component on small scales by
the high-$\ell$ CMB, LSS, and Lyman-$\alpha$ data in Section
\ref{smallscales}.   In Section \ref{models} we explore the parameter
space of hybrid inflation and double D-term inflation 
models which could give parameters suggested
by the previous phenomenological analysis.
We compare the $k^3$ spike to more general 
$k^n$ spikes in Section \ref{knspike}.
We give conclusions in Section \ref{conclusions}.  In Appendix \ref{appendixa} we
clarify the relation between the $k^3$ spectrum (in modern notation) 
and the causality prediction made in the early literature.

\section{Motivation for the $k^3$ component}
\label{motivation}

Prior to the idea of inflation, it was argued on the basis of
causality that any fluctuations which are created independently of
each other must have a correlation function that vanishes  at least
as fast as $k^3$ as $k\to 0$.  This was first proven in the context
of Newtonian gravity and then extended to general relativity in 
Ref.\ \cite{Abbott-Traschen}.  In Appendix \ref{appendixa} we
recapitulate (and translate into modern notation) this argument.
The essential observation is that a density perturbation 
$\delta\rho(\mf x)/\rho$ which consists of contributions from causal
processes originating at positions $\mf x_a$ has the form
\be
  {\delta\rho(\mf x)\over\rho} = \sum_a F_a(\mf x - \mf x_a),
\ee
where causality demands that the monopole and dipole moments of 
$F_a$ vanish in a volume extending over distances greater than
the causal horizon \cite{Traschen, Traschen:1983ee}.  In Fourier space, this implies
that $F_{a\mf k}$ must vanish at least as fast as $k^2$ for small
$k$.  Therefore the correlation $\langle |\delta\rho_{\mf 
k}/\rho|^2\rangle$ falls like $k^4$ at small $k$.  
Appendix \ref{appendixa} shows how this corresponds to a $k^3$
spectrum for the curvature perturbation.  

Inflationary perturbations evade this constraint by having been in
causal contact with each other before being driven out of causal
contact by inflation.  But perturbations produced at the end of
inflation, or between two consecutive stages of inflation,  for
example during a phase of tachyonic preheating, should obey the
constraint.   A $k^3$ component could lead to interesting effects at
short scales, such as the production of primordial black holes.  It
could conceivably also distort the spectrum of CMB fluctuations at
longer scales, if it is appropriately cut off at the short scales.  
Of course, there is no reason to believe that the $k^3$ growth should
continue to arbitrarily large $k$, since it is derived in the region
$k \to 0$, and unrestricted growth at large $k$ would lead to an
unphysical UV divergence in the power.  The value of the cutoff $k_{\rm c}$
beyond which the effect vanishes depends on the particular model
giving rise to it.  In the case of tachyonic preheating,  the cutoff
is determined by the mass  parameter ($m^2_\sigma$) of the tachyonic
field, since only modes with physical wave number $k^2 < a^2 |m^2_\sigma|$
at the time of preheating undergo exponential growth. 

Explicit examples of $k^3$ spectral components have been discussed in
Refs.\ \cite{LLMW, FK, STBK}; they were generated by preheating at the end of
chaotic inflation.  It was shown that those models do not have enough
freedom to give an observably large $k^3$ component once the COBE
normalization is imposed.  On the other hand, hybrid inflation models
have more free parameters and thus have more likelihood to produce an
observably large effect.  In Ref.\ \cite{julien1} it was noted that
double (D-term) hybrid inflation models \cite{sak, julien2, kanazawa}
can produce a spike with a $k^3$ spectrum and a cutoff, due to a
tachyonic instability which triggers the transition between the two
stages of inflation.  More recently it has been shown 
\cite{k3, k3_2} that tachyonic preheating even in the simplest model
of hybrid inflation can generate an observably
large $k^3$ component or alternatively large non-Gaussianity (see also
Refs.\ \cite{Enqvist-etal, Enqvist:2004ey, Enqvist:2005qu, Enqvist:2005nc}) for certain ranges of parameters.   This effect
arises at second order in the cosmological perturbation, and its
importance is controlled by how large  the fluctuations of the
tachyon field can become before their back-reaction ends inflation.

In Section \ref{models} we will make a detailed study of the
predictions of tachyonic preheating in the hybrid inflation for the
$k^3$ perturbation and how it compares to the CMB and
LSS data.  In the next section we describe these
data.

\section{Data}
\label{data}

To explore the experimental evidence for the $k^3$ component, we use
CosmoMC, a publicly available Markov-chain Monte-Carlo (MCMC) engine for
exploring cosmological parameter space \cite{cosmomc}.\footnote{See
http://www.cosmologist.info/cosmomc.}  We use the first year and
three-year WMAP data (WMAP1 and WMAP3)  \cite{WMAP1-1, WMAP3-1} as primary data sets,
analyzing them with the July 2005 and May 2006 versions of CosmoMC,
respectively. For completeness and as a consistency check, we consider
both WMAP1 and WMAP3 in this work.

We include the high-$\ell$ CMB data, CBI \cite{CBI} and ACBAR
\cite{ACBAR}, to explore the evidence for the $k^3$ component on
small scales. The ranges of data we use are $300<\ell<3500$ (14
data points) for CBI and $300<\ell<3000$ (13 data points) for
ACBAR.\footnote{The offset lognormal matrix for 2000 + 2001 CBI
was incorrectly incorporated into May 2006 or earlier versions of
CosmoMC \cite{cbimatrix}, so we use our offset lognormal matrix in
this paper. See Ref.\ \cite{cbimatrix} for a discussion on the impact
of using the wrong matrix; also see the CBI website,
http://www.astro.caltech.edu/\~{}tjp/CBI/data2004, for a link to the
discussion on this issue.} The first bands of both data are not
included in the analysis since they are not well constrained.

The LSS data, 2dFGRS \cite{2dF, 2dF2005} and SDSS main galaxy sample
\cite{SDSS},\footnote{2002 and 2005 2dFGRS data are used along with
WMAP1 and WMAP3, respectively; the differences of the results,
however, are not significant.} are also used in our analysis. To use
the linear theory, we set $k_{\rm max} = 0.15h\ {\rm Mpc}^{-1}$ in
the galaxy-galaxy power spectrum as recommended by 2dFGRS and SDSS;
their ranges are $0.022\ {\rm Mpc}^{-1} < k/h < 0.147\ {\rm Mpc}^{-1}$ for 2dFGRS (32
data points) and $0.016\ {\rm Mpc}^{-1} < k/h < 0.154\ {\rm Mpc}^{-1}$ for SDSS (17
data points). SDSS also probes the nonlinear regime, which allows us
to explore the possibility of fitting the nonlinear regime with the
$k^3$ component. The ranges are extended to $k/h = 0.205\ {\rm
Mpc}^{-1}$ (2 more data points). Further discussions can be found in Section \ref{nonlinear}.
We also include the SDSS Luminous Red Galaxies (LRG) power spectrum
\cite{Tegmark:2006az}; the range is $0.012\ {\rm Mpc}^{-1} < k/h 
< 0.087\ {\rm Mpc}^{-1}$ (14 data points) for linear theory, and it is 
extended to $k/h = 0.203\ {\rm Mpc}^{-1}$ (6 more data points) for nonlinear 
theory. The results of 2003 \cite{SDSS} and 2006 SDSS \cite{Tegmark:2006az} 
data are referred to as SDSS1 and SDSS4, respectively, since they used the 
SDSS data release 1 and 4, respectively.

Lyman-$\alpha$ forest data allow us to explore the high-$k$ regime.
We use two different Lyman-$\alpha$ data sets to test the $k^3$
component. Viel {\it et al}.\ \cite{Viel} used the LUQAS sample
\cite{Kim}, which has $z = 2.125$, and its range is $0.0034\ ({\rm km/s})^{-1} < k <
0.027\ ({\rm km/s})^{-1}$; they also reanalyzed the result of Croft
{\it et al}.\ \cite{Croft} ($z=2.72$) in the same range. The SDSS
Lyman-$\alpha$ data \cite{sdsslya1, sdsslya2} have $2<z<4$ and the
range is $0.0013\ ({\rm km/s})^{-1} < k < 0.02\ ({\rm km/s})^{-1}$.\footnote{For the
SDSS Lyman-$\alpha$ data, we do not use the default code in CosmoMC;
instead, we use the patch provided by An\v{z}e Slosar. See
http://www.slosar.com/aslosar/lya.html.} The data sets we use, from
CMB, LSS, to Lyman-$\alpha$ forest, cover $k$ space from
$10^{-4}\ {\rm Mpc}^{-1}$ to 3 Mpc$^{-1}$ and provide full sensitivity to a possible
$k^3$ component on observable scales.

In our comparison of  the power-law $\Lambda$CDM model
with the $k^3$ and $k^3_{\rm c}$ models, 
we do not include the BB polarization power spectrum or the
Sunyaev-Zel'dovich (SZ) effect. 
Most settings in CosmoMC are by default. 
Typical MCMC chains have on the order of $10^5$ points and we search these
chains to find the best fit points.

\section{Evidence from WMAP: Large Scales}
\label{fittingwmap}

In this section we focus on the large scale (small $k$) region  of
the spectrum, using primarily  the WMAP data to fit the $k^3_{\rm c}$
model; for comparison we will also show the effect of combining with
other CMB (CBI and ACBAR), LSS (2dFGRS or SDSS), or Lyman-$\alpha$
data.
To get a significant effect at low $k$, the cutoff $k_{\rm c}$ is
necessary in order to avoid being dominated by the large amount of extra
power at high $k$.
Table \ref{tab1} lists the best fit $k^3_{\rm c}$ models for
WMAP alone, and in combination with other CMB (CBI and ACBAR), LSS
(2dFGRS or SDSS), or Lyman-$\alpha$ data. 

\begin{table}[htp]
\caption{Best fit $k^3_{\rm c}$ models of fitting WMAP alone, 
and in combination with other CMB (CBI and ACBAR), LSS
(2dFGRS or SDSS), or Lyman-$\alpha$ data. ${\chi^2}^\prime$ is the contribution to $\chi^2$
from all data excluding WMAP.}
\label{tab1}
\renewcommand{\arraystretch}{1.5}
\vspace{2ex}
\centering{\begin{tabular}{ccccccc}
\hline\hline
\raisebox{-1.7ex}[0pt][0pt]{Data} & $\chi^2/{\chi^2}^\prime$ & $\Delta \chi^2/\Delta{\chi^2}^\prime$ & $n_s$ & \raisebox{-1.7ex}[0pt][0pt]{$r_3(k_0)$} & $k_{\rm c}$ & $r_3$\\
& ($\Lambda$CDM) & ($k^3_{\rm c}$) & $(k_0)$ & & $({\rm Mpc}^{-1})$ & $(k_{\rm c})$\\
\hline
WMAP1 & 1428.8/-- & 3.6/-- & 1.005 & 0.0911 & 0.00394 & 0.694\\
+ CMB & 1463.1/30.9 & 3.2/$-2.5$ & 0.987 & 0.0782 & 0.00385 & 0.561\\
+ 2dFGRS & 1463.2/34.4 & 1.7/0.0 & 1.010 & 0.0280 & 0.00408 & 0.236\\
+ SDSS1 (linear) & 1445.9/16.9 & 0.5/$-0.2$ & 0.987 & 0.0715 & 0.00356 & 0.407\\
+ SDSS4 (linear) & 1443.3/12.5 & 3.8/$-0.4$ & 0.983 & 0.0123 & 0.00364 & 0.747\\
\hline
WMAP3 & 11252.3/-- & 5.4/-- & 0.944 & $1.62\times10^{-5}$ & 0.0388 & 0.139\\
+ CMB & 11285.9/33.3 & 4.1/0.4 & 0.949 & $2.38\times10^{-5}$ & 0.0373 & 0.178\\
+ 2dFGRS & 11290.8/38.4 & 3.7/$-0.2$ & 0.947 & $1.08\times10^{-5}$ & 0.0383 & 0.0890\\
+ SDSS1 (linear) & 11274.0/18.9 & 4.5/$-0.1$ & 0.952 & $1.26\times10^{-5}$ & 0.0396 & 0.113\\
+ SDSS4 (linear) & 11264.9/12.2 & 1.1/$-0.8$ & 0.955 & $1.12\times10^{-5}$ & 0.0347 & 0.0666\\
\hline
\raisebox{-1.7ex}[0pt][0pt]{SDSS1 +} WMAP1 & 1452.5/23.6 & 0.9/2.2 & 0.959 & $8.36\times10^{-7}$ & 0.147 & 0.397\\
\hskip10ex WMAP3 & 11278.1/24.3 & 0.9/2.2 & 0.943 & $6.18\times10^{-7}$ & 0.131 & 0.222\\
\hline
\raisebox{-1.7ex}[0pt][0pt]{SDSS4 +} WMAP1 & 1453.7/24.3 & 1.6/4.1 & 0.950 & $24.0\times10^{-7}$ & 0.0850 & 0.223\\
\hskip10ex WMAP3 & 11276.9/24.2 & 0.4/2.4 & 0.945 & $9.76\times10^{-7}$ & 0.0725 & 0.0567\\
\hline
\raisebox{-1.7ex}[0pt][0pt]{Ly-$\alpha$ +} WMAP1 & 1454.1/24.9 & 2.5/3.3 & 0.973 & $1.05\times10^{-9}$ & 1.49 & 0.521\\
\hskip8ex WMAP3 & 11279.9/26.9& 2.5/3.8 & 0.959 & $0.612\times10^{-9}$ & 1.49 & 0.331\\
\hline\hline
\end{tabular}}
\end{table}

\begin{figure}[ht]
\centering{\includegraphics[width=\textwidth]{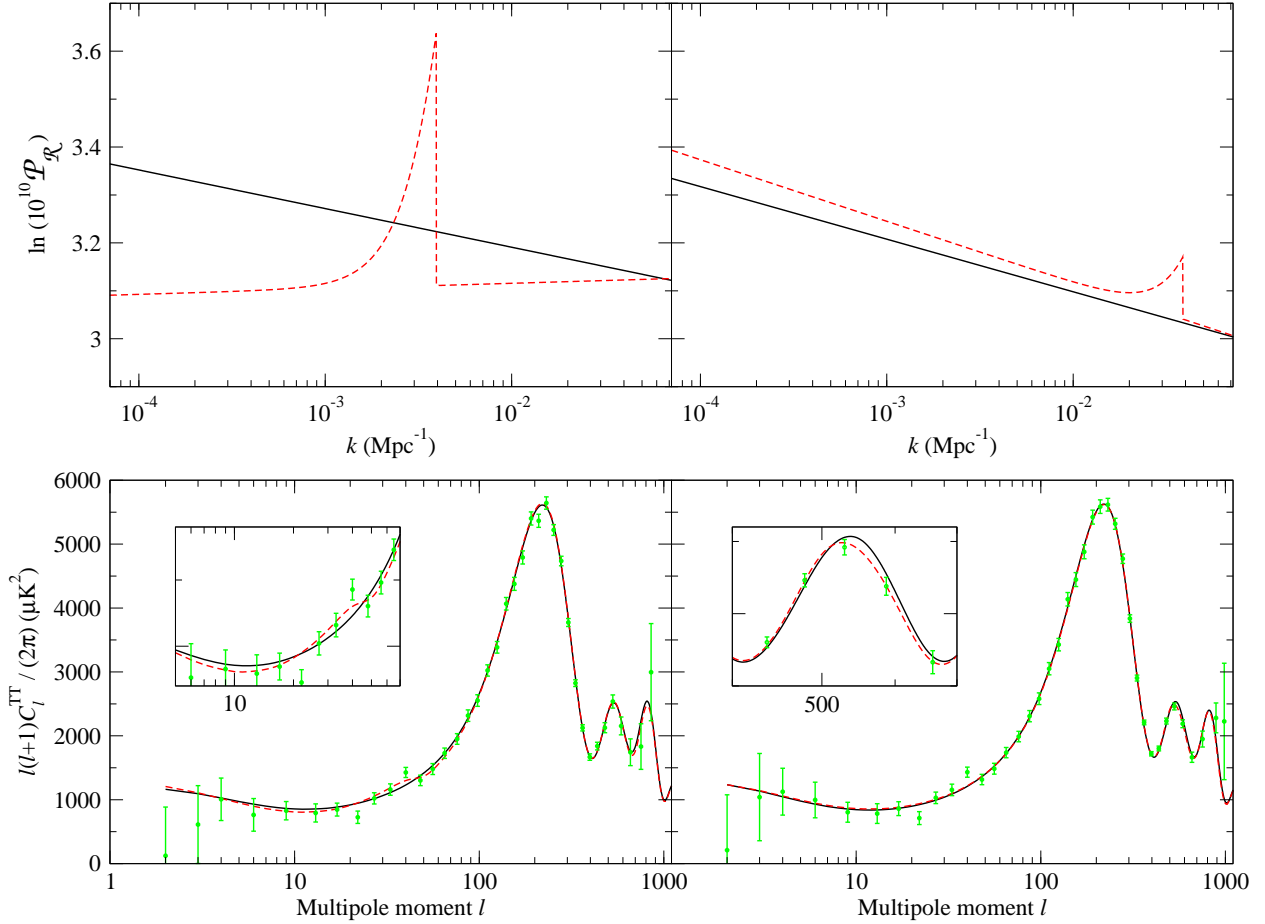}}
\caption{The best fit primordial power spectra (upper) and $C_\ell^{\rm TT}$ spectra (lower) for the power-law 
$\Lambda$CDM model (solid, black) and the $k^3_{\rm c}$ model (dashed, red) using the WMAP1 (left) and
WMAP3 (right) data (dotted, green). The scales of the wave numbers $k$ (Mpc$^{-1}$) are shifted to match those of the multipole moment $\ell$.}
\label{w1w3}
\end{figure}

As can be seen in Table \ref{tab1}, the $k_{\rm c}^3$ model improves the fit
to the data, relative to that of the power-law $\Lambda$CDM model, 
by a reduction in the $\chi^2$ of 
$\Delta \chi^2=3.6$ for the first year WMAP data, where
\be
	\Delta\chi^2 = \chi^2(\Lambda{\rm CDM}) - \chi^2(k^3_{\rm c}).
\ee
Recall that $\Delta \chi^2=4.7$ for the running spectral index +
tensor model \cite{msc}, and so the $k^3_{\rm c}$ model is not as good as
the running spectral index + tensor model when the first year WMAP
data are used (both models have eight parameters). Nevertheless, the
$k^3_{\rm c}$ model gives a large amplitude ratio
($r_3(k_{\rm c})=0.69$), and hence WMAP1 allows for a relatively 
large extra component in addition to 
the nearly scale-invariant spectrum, as shown in the top-left panel of \fig{w1w3}. More
intriguingly,  for the three-year WMAP data, we find an improvement
of $\Delta \chi^2=5.4$; recall that $\Delta \chi^2=3.1$ for the
running spectral index + tensor model \cite{msc}, and so the $k^3_{\rm
c}$ model gives a better fit than the running spectral index + tensor
model using WMAP3. The top-right panel of \fig{w1w3} shows the spectra; it is interesting
that a small modification ($r_3(k_{\rm c})=0.14$) to the nearly
scale-invariant spectrum can give such a large $\Delta
\chi^2$.

To better understand why the $k^3$ term improves the fits, the
$C_\ell^{\rm TT}$ spectra are plotted. The lower part of \fig{w1w3} shows the best
fit $C_\ell^{\rm TT}$ spectra for the power-law $\Lambda$CDM model
(solid, black) and the $k^3_{\rm c}$ model (dashed, red) using the
WMAP1 (left) and WMAP3 (right) data (dotted, green); the
$C_\ell^{\rm TE}$ spectra are hardly changed, so they are not shown.
It can be seen that both spectra are consistent over a large range,
but the $C_\ell^{\rm TT}$ spectrum of the $k^3_{\rm c}$ model
provides a better fit at low multipoles ($\ell \sim10 - 50$) when WMAP1
is used. Therefore, the $k^3_{\rm c}$ model can help explain the 
``glitch'' in the low multipoles of the $C_\ell^{\rm TT}$
spectrum. It has also been argued that an exponentially increasing
step \cite{pkrecon, pkrecon2} or oscillations \cite{future,
Hamann:2007pa} in the primordial power spectrum fit the low
$C_\ell^{\rm TT}$ multipoles better than the nearly scale-invariant
spectrum. Since the peak of the $k^3_{\rm c}$ model is similar to an
exponentially increasing step or oscillations to some extent, it
is not surprising that the $k^3$ model can offer
an alternative physical explanation for the improved fits found
in Refs.\ \cite{pkrecon, pkrecon2, future, Hamann:2007pa}.

As for WMAP3, Table \ref{tab1} and \fig{w1w3} show that the best fit
$k^3$ peak appears around 0.04 Mpc$^{-1}$, close to
the second peak ($\ell \sim 540$) of the $C_\ell^{\rm TT}$
spectrum (see the right panels of \fig{w1w3}). This can be understood
due to the failure of the best fit $\Lambda$CDM model to match
the data within $1\sigma$ in this region.  Due to improved sensitivity
of WMAP3 to these higher multipoles, the ``glitch'' at $\ell \sim 540$
is statistically more significant to WMAP3 than to WMAP1, where
the $\ell \sim 10 - 50$ glitch took precedence.  
This explains why the best fit power spectra of WMAP1 and WMAP3
are very different (see \fig{w1w3}). We emphasize that these 
results  are consistent with each other. One could have
the $k^3$ peak appear at 0.004 Mpc$^{-1}$ for WMAP3, but the
resultant $\Delta\chi^2$ is smaller ($\Delta\chi^2=2.0$); similarly,
one could adjust the $k^3$ component to fit the high multipoles of
WMAP1, but the $\Delta\chi^2=1.2$ is smaller than that of fitting the
low multipoles.
(Of course, it is possible for a single data set to favor, to some
extent, the simultaneous appearance of the $k^3$ feature at different scales; one of the objective
of this paper is to constrain such an effect, which will be given in
Section \ref{constraint}.)
Obviously, a more complicated primordial power
spectrum could have a better fit. In particular, introducing an
additional $k^3$ peak at 0.004 Mpc$^{-1}$ while keeping the $k^3$ peak at
0.04 Mpc$^{-1}$ fits better than either peak by itself.
References \cite{future, Hamann:2007pa} make a similar
observation; by changing the location and amplitude of the
oscillations, their model is able to fit different ranges of
multipoles.
Unlike the oscillations which affect all multipoles beyond the second peak of the CMB power spectrum, the effect of the $k^3$ component for WMAP3 in the best fit model is localized at the second peak.

\begin{figure}[ht]
\centering{\includegraphics[width=5in]{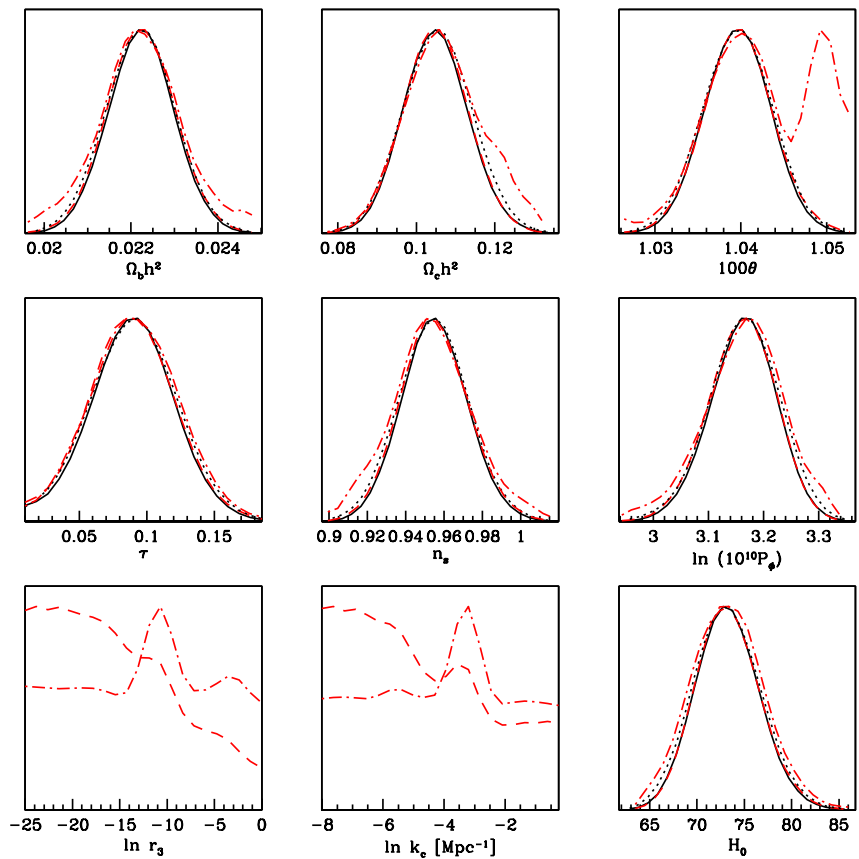}}
\caption{The distributions of the parameters for the power-law $\Lambda$CDM model (black) and the $k^3_{\rm c}$
model (red) using the WMAP3 data. Solid ($\Lambda$CDM) and dashed ($k^3_{\rm c}$) lines are
marginalized probabilities; dotted ($\Lambda$CDM) and dot-dashed ($k^3_{\rm c}$) lines are mean
likelihoods of samples. ($k_0=0.002\ {\rm Mpc}^{-1}$.)}
\label{w3_k3}
\end{figure}

\Fig{w3_k3} shows the distributions of the parameters for the
power-law $\Lambda$CDM model (black) and the $k^3_{\rm c}$ model (red) using the WMAP3 data. Solid ($\Lambda$CDM) and dashed ($k^3_{\rm c}$)
lines are marginalized probabilities, and dotted ($\Lambda$CDM) and dot-dashed ($k^3_{\rm c}$) lines are mean
likelihoods of samples; for Gaussian distributions, they should be
the same. The mean likelihoods indicate whether the parameters
are really being constrained, or affected by the priors and the
volume of samples; they also show how good a fit one can expect
\cite{cosmomc}. There are four Markov chains for the model, and they
satisfy the Gelman and Rubin convergence test $R-1<0.03$. The Gelman
and Rubin ``variance of chain means" / ``mean of chain variances"
convergence test generally demands that $R-1<0.1$ for each parameter; for
example, the WMAP collaboration demands that $R-1<0.1$ \cite{WMAP3-1}.
However, smaller numbers usually indicate better 
convergence.\footnote{See CosmoMC's website, 
http://www.cosmologist.info/cosmomc,
for further discussions.}

One can see from \fig{w3_k3} that introducing the $k^3$ component does not have significant statistical influence on the standard parameters. Nevertheless,
there are inconsistencies between the
marginalized probabilities and the mean likelihoods of samples at the
higher ends of $\Omega_bh^2$, $\Omega_ch^2$, $\theta$, $n_s$, and $\ln
P_\phi$. These inconsistencies, along with the small bumps of $\ln
r_3$ and $\ln k_{\rm c}$, indicate that there is another local
best fit point in the parameter space, i.e.\ the
$k^3_{\rm c}$ model, which fits the low multipoles.  This can be
seen in the corresponding probabilities for WMAP1 (which we do not
show): there the smaller bumps seen in the distributions for WMAP3
become the more significant features,
and shift the most likely values.

We also notice that the marginalized probabilities of $\ln r_3$ and $\ln k_{\rm c}$ are far from Gaussian. This is related to the priors of $\ln r_3$ and $\ln k_{\rm c}$:
\bea
	&-25 < \ln r_3 (k_0)< 0,&\\
	&-8 < \ln k_{\rm c}\ [{\rm Mpc}^{-1}] < -0.25.&
\eea
To justify these priors, we note that the amplitude ratio $\ln
r_3 (k_0)=-25$ gives a negligible $k^3$ component even on small
scales, i.e.\ $r_3(1\ {\rm Mpc}^{-1}) \sim 10^{-3}$.
On the other hand, if $\ln r_3 (k_0) > 0$, then the $k^3$
component will dominate over the primordial power spectrum and it
should have been observed. The prior of $\ln k_{\rm c}$ is set to
cover observable $k$ space: $\ln k\ [{\rm Mpc}^{-1}] = -8$ roughly
corresponds to the largest observable scale and $\ln k\ [{\rm
Mpc}^{-1}] = -0.25$ is the highest value found to be called by
CosmoMC when the CMB data are used. These are not true priors in
any rigorous sense. From a phenomenological perspective these models
are unbounded from below and any proper Bayesian estimate of parameters
is likely to be swamped by the large part of the allowed parameter space that 
provides effectively no observed deviations from scale invariance. This can
be seen in the marginalized probabilities of \fig{w3_k3}. However,
for understanding to what extent the data allow such deviations these priors
permit practical exploration of the allowed parameter space. 

The choice of priors still allows for degeneracies. 
For any value of $\ln r_3$ in the prior, taking
the minimum value of $\ln k_{\rm c}$ makes the $k^3$
contributions negligible, and vice versa. This fact explains why the
higher tails  of the $\ln r_3$ and $\ln k_{\rm c}$
distributions in \fig{w3_k3} do not decay,  since the $k^3_{\rm c}$
model becomes the power-law $\Lambda$CDM model in that region and has
a large volume of samples.  Similarly, the lower ends of the $\ln
r_3$ and $\ln k_{\rm c}$ distributions have a larger range where
the $k^3$ component is negligible and hence give rise to higher
values of marginalized probabilities, as seen in \fig{w3_k3}.

\begin{figure}[htp]
\centerline{\includegraphics[width=3.5in]{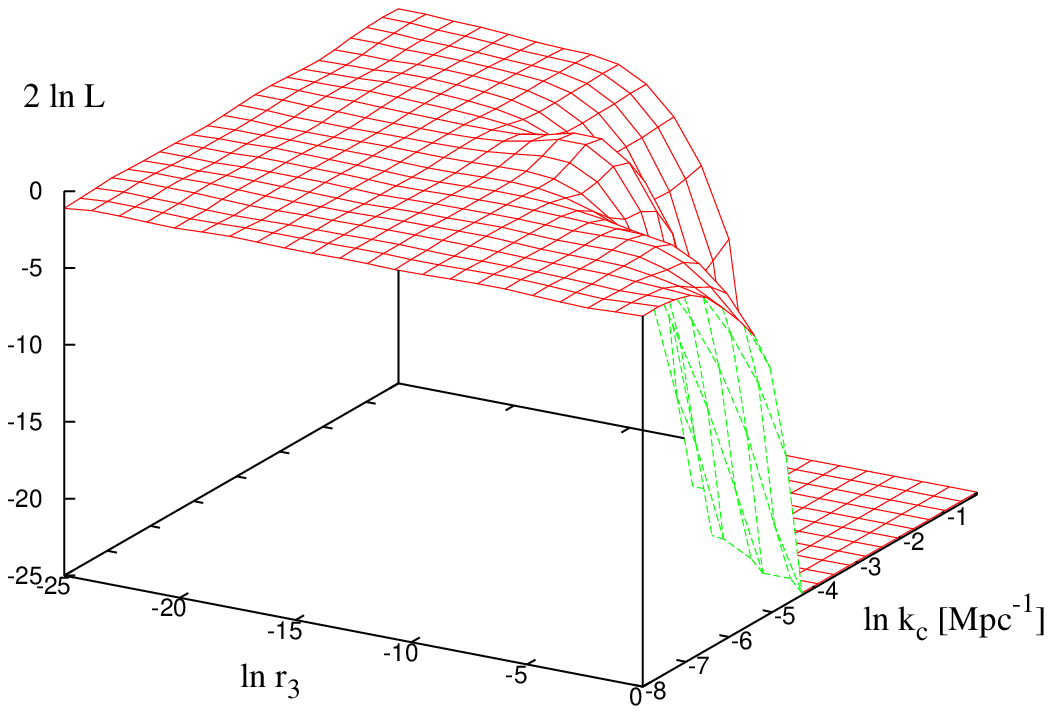}\includegraphics[width=3.5in]{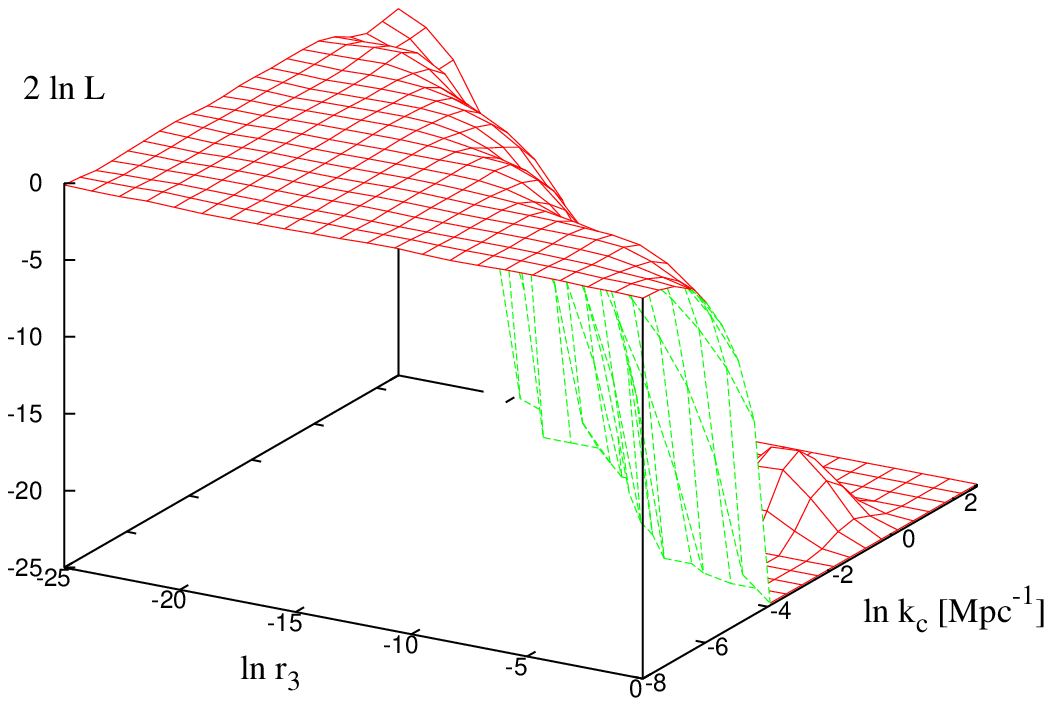}}
\centerline{\includegraphics[width=3.5in]{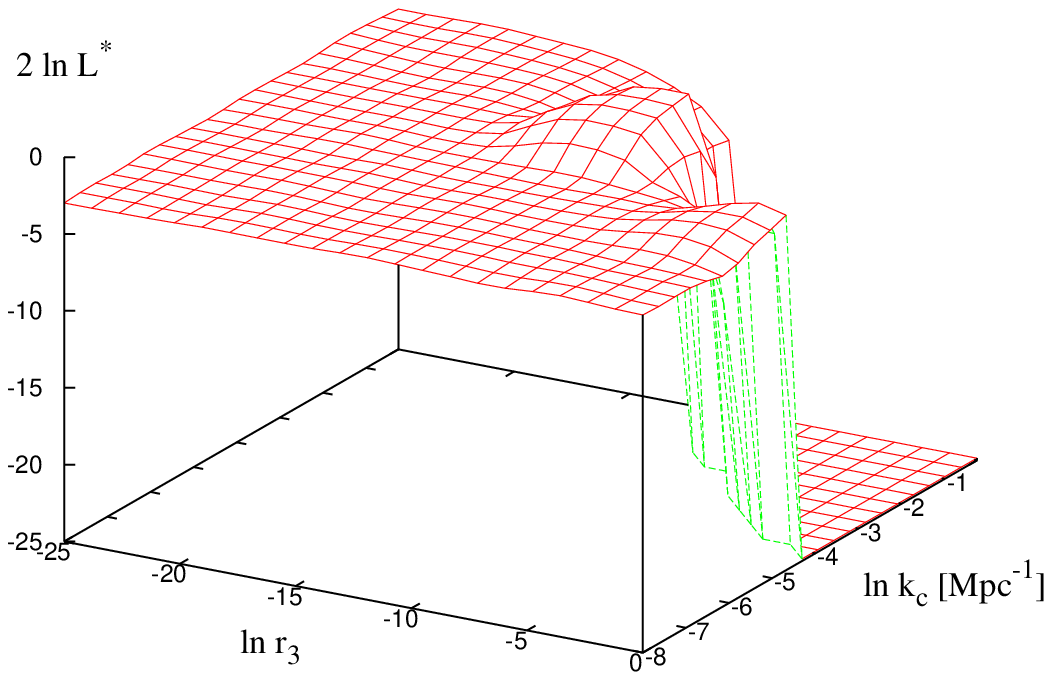}\includegraphics[width=3.5in]{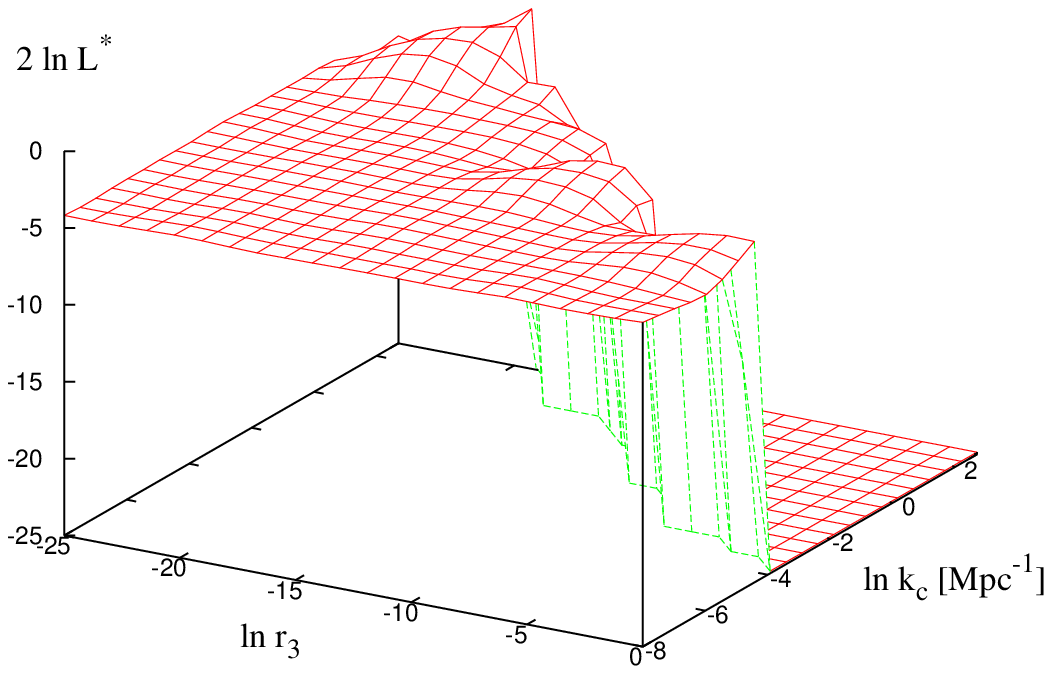}}
\centerline{\includegraphics[width=3.2in]{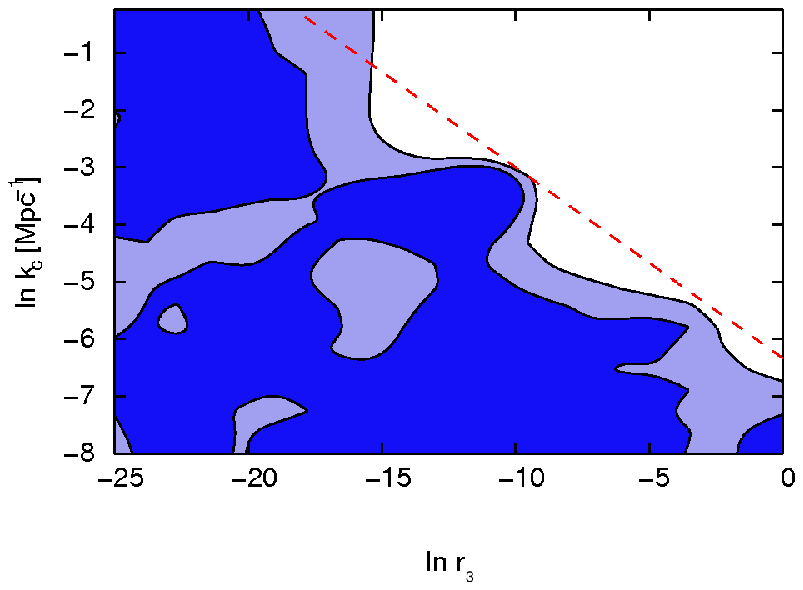}\includegraphics[width=3.2in]{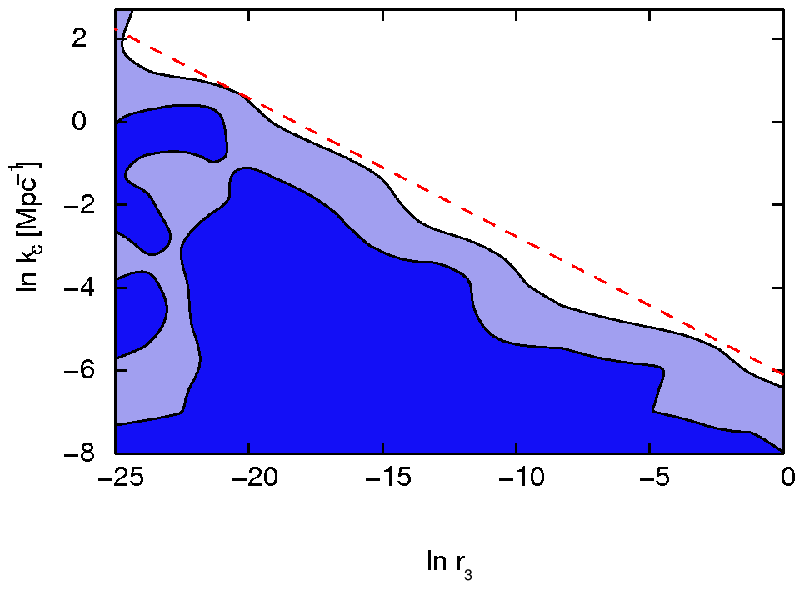}}
\caption{The two-dimensional marginalized probabilities (top; $2 \ln L$), mean likelihoods of samples (middle; $2\ln L^*$), and marginalized probability contours (bottom; 68\% and 95\% CL) of $\ln r_3$ and $\ln k_{\rm c}$ for the $k^3_{\rm c}$ model using the WMAP3 alone (left), and in combination with SDSS4 (linear) and Viel {\it et al}.\ Lyman-$\alpha$ data (right). The dashed (red) lines in the marginalized probability contours give the best bounds on the amplitude ratio over the observable wave numbers (see Section \ref{constraint}).
The surface plots have an artificial floor of $2\ln L=-25$ inserted for plotting
purposes (see footnote \ref{likelihoodeps}).}
\label{w3_k3_problike}
\end{figure}

Plotting the two-dimensional marginalized surfaces makes the argument
clearer. \Fig{w3_k3_problike} shows the two-dimensional
marginalized probabilities (and their 68\% and 95\% CL contours) and
mean likelihoods of samples of $\ln r_3$ and $\ln k_{\rm c}$ for
the $k^3_{\rm c}$ model using the WMAP3 alone (left), and in combination with SDSS4 (linear) and Viel {\it et al}.\ Lyman-$\alpha$ data (right). (See Section \ref{constraint} for further discussions.)
The waterfall shape 
surfaces demonstrate that the high $\ln r_3$ and $\ln k_{\rm c}$
regions are ruled out by the data. The high plateaus are the regions
having a negligible $k^3$ component and a large volume of samples. The
peaks in the probabilities and mean likelihoods of samples represent the
best fit models.
We normalize the marginalized probabilities and mean likelihoods of samples at their maximum points and show them in logarithmic scales, so that one can see how far the plateaus are from the best fit points in terms of $\Delta\chi^2 \sim -2\ln L$.\footnote{\label{likelihoodeps}Some of the points at the bottom of the waterfall are zero, so their logarithm will give negative infinities. To solve this problem, we plot $2\ln (L+\eps)$ instead, where we take $\eps=4\times10^{-6}$ so that the minimum values are $-25$, roughly $5\sigma$ from the best fit points.}
As we discussed above, there are two peaks in the
mean likelihood of samples, the smaller one corresponding to fitting
the glitch at low multipoles.\footnote{The peaks are not obvious, since they are plotted on a logarithmic scale.} Since the plateau is near the 1$\sigma$
confidence level, the projection of the two-dimensional marginalized
probabilities shows many contours where the surface crosses
$1\sigma$; see the bottom panels of \fig{w3_k3_problike}.

Table \ref{tab2} lists the best fit models and marginalized values
for the power-law $\Lambda$CDM model and the $k^3_{\rm c}$ model using the WMAP3 data. Due to
the points just discussed, the one-dimensional marginalized values do not
give a good estimate or reliable upper limits on $\ln r_3$ and
$\ln k_{\rm c}$. (However, we will give constraints on the maximum
amplitude ratio in Section \ref{constraint}.) Nevertheless,
the improvements of fitting the irregularities at the low multipoles
($\ell \sim 10 - 50$) or the second peak ($\ell \sim 540$) could give
an additional contribution as large as 14\% to 69\% to the nearly scale-invariant spectrum
(see Table \ref{tab1}). This observation not only provides evidence
for the $k^3_{\rm c}$ model but also demonstrates the extent to
which the data still leave open the possibility of large
deviations from scale invariance in the power spectrum.

Furthermore, \fig{w3_k3} shows that the best fit points can have significant
variations from the $\Lambda$CDM values, as can be seen from the mean
likelihoods of samples. With our choice of priors these points 
contribute little to the marginalized likelihoods, but it is surprising that
adding localized features can move physical parameters such as the angular
size of the sound horizon by order ``1$\sigma$.'' With future polarization
data this should be less of an issue, but the robustness of physical
parameters to the physics of inflation could be important for future
dark energy studies that rely on CMB-derived physical parameters.

\begin{table}[htp]
\caption{The best fit power-law $\Lambda$CDM model and the $k^3_{\rm c}$ model, and their marginalized (mean, 68\% CL) values for the WMAP3 data.}
\label{tab2}
\renewcommand{\arraystretch}{1.5}
\vspace{2ex}
\centering{\begin{tabular}{ccccc}
\hline\hline
\raisebox{-1.7ex}[0pt][0pt]{Parameter} & \multicolumn{2}{c}{$\Lambda$CDM} & \multicolumn{2}{c}{$k^3_{\rm c}$}\\
& Best fit & Mean & Best fit & Mean\\
\hline
$10^2\Omega_bh^2$ & 2.22 & $2.22_{-0.07}^{+0.07}$ & 2.24 & $2.22_{-0.07}^{+0.07}$\\
$\Omega_ch^2$ & 0.105 & $0.105_{-0.008}^{+0.008}$ & 0.112 & $0.105_{-0.008}^{+0.008}$\\
$10^2\theta$ & 1.040 & $1.040_{-0.004}^{+0.004}$ & 1.043 & $1.040_{-0.004}^{+0.004}$\\
$\tau$ & 0.090 & $0.090_{-0.029}^{+0.029}$ & 0.092 & $0.089_{-0.030}^{+0.030}$\\
$n_s(k_0)$ & 0.952 & $0.955_{-0.016}^{+0.016}$ & 0.944 & $0.955_{-0.017}^{+0.017}$\\
$\ln 10^{10} {\cal P_\phi}(k_0)$ & 3.17 & $3.16_{-0.06}^{+0.06}$ & 3.21 & $3.16_{-0.06}^{+0.06}$\\
$\ln r_3(k_0)$ & -- & -- & $-11.0$ & $-15.3^a$\\
$\ln k_{\rm c}$ [Mpc$^{-1}$] & -- & -- & $-3.25$ & $-4.81^a$\\
$h$ & 0.728 & $0.732_{-0.032}^{+0.032}$ & 0.716 & $0.733_{-0.032}^{+0.032}$\\
\hline\hline
\multicolumn{5}{l}{\footnotesize\hskip7.5pt$^a$See Section \ref{constraint} for the constraints.}\\
\end{tabular}}
\end{table}

\section{Fitting Small Scales}
\label{smallscales}

To explore the evidence for the $k^3$ component on small scales, we
now focus on other CMB, LSS, and Lyman-$\alpha$ data. The results are
discussed in detail in the following subsections.

\subsection{Fitting High-$\ell$ CMB or LSS Data}

Including the CBI and ACBAR data allows us to test for the possible
presence 
of the $k^3$ component in the high-$\ell$ regime. The best fit models
are listed in Table \ref{tab1}. As can be been, the best fit power
spectra of joint WMAP, CBI, and ACBAR are close to those of
fitting WMAP alone. This implies that the improvements to the fits
come primarily from the WMAP data ($\ell \sim 10 - 50$ or $\ell \sim 540$),
and hence the CBI and ACBAR data do not seem to give any additional
evidence for a $k^3$ component.  Although there is some excess
power in the CBI and ACBAR data at small scales, $\ell > 1000$,
this cannot be attributed to the $k^3$ effect because Silk damping
(also known as diffusion damping)
kills any sensitivity of the CMB to the primordial power spectrum
in this region.  Any perturbation to the 
 exponential damping tail seen in the CMB angular power
spectrum can only arise from astrophysical effects long after 
inflation.  Thus, no evidence for a $k^3$ component at large
$k$ can be obtained just using CMB data.

We therefore consider the evidence for a $k^3$ component at large $k$
using LSS data. Table \ref{tab1} shows that
the locations and amplitudes of the $k^3$ peaks are close to those
inferred from the WMAP  data alone. We see  that adding a
$k^3$ peak at large $k$ ($<0.15h\ {\rm Mpc}^{-1}$ for 2dFGRS and SDSS1 [linear]; $<0.1h\ {\rm Mpc}^{-1}$ for SDSS4 [linear]) fails to improve
the fit to the galaxy-galaxy power spectrum.

To further quantify the significance of the WMAP data, we compute a
quantity ${\chi^2}^\prime$, defined to be  the contribution to the total
$\chi^2$ coming from all data except WMAP.  These values are shown
in Table \ref{tab1}. Since the $k^3$ peaks at $\sim 0.004$
Mpc$^{-1}$ (WMAP1) are outside the ranges of the CBI, ACBAR, 2dFGRS,
and SDSS (linear) data which we use (see Section \ref{data}), these data are not
directly sensitive to such a $k^3$ component.  Indirectly, however,
they have an effect, since the best fit $k^3_{\rm c}$ model of WMAP1
favors a bluer tilt  than do the CBI and ACBAR data.  This explains
why the $k^3$ peak at low $k$ is somewhat disfavored by these data:
$\Delta{\chi^2}^\prime=-2.5$.  On the other hand, for the  $k^3$ peak
at $\sim 0.04$ Mpc$^{-1}$ (WMAP3), the high-$\ell$ CMB and LSS data exhibit no
preference, giving $\Delta{\chi^2}^\prime$ of order $O(0.5)$. Thus,
although the CMB and LSS data do not give positive evidence, they are
still consistent with a $k^3$ component which introduces a
perturbation on the nearly scale-invariant primordial spectrum whose
magnitude is  7\% to 18\%.

\subsection{Fitting the Nonlinear Regime}
\label{nonlinear}

The SDSS data also probe the nonlinear regime, and we include data 
up to $k/h = 0.2\ {\rm Mpc}^{-1}$, as recommended by the SDSS team.
As reference models, we consider the power-law $\Lambda$CDM model taking into
account nonlinear evolution effects; we also fit the $k^3_{\rm c}$ model using the nonlinear theory.
The total $\chi^2$'s, the $\chi^2$'s of the galaxy-galaxy power
spectra, and the best fit $k^3_{\rm c}$ models are shown in Table
\ref{tab1}. As can be seen from the table, the results of WMAP1 or WMAP3 plus SDSS1 are consistent, and both $k^3_{\rm c}$ models fit slightly better than the
power-law $\Lambda$CDM model.

\begin{figure}[ht]
\centering{\includegraphics[width=4in]{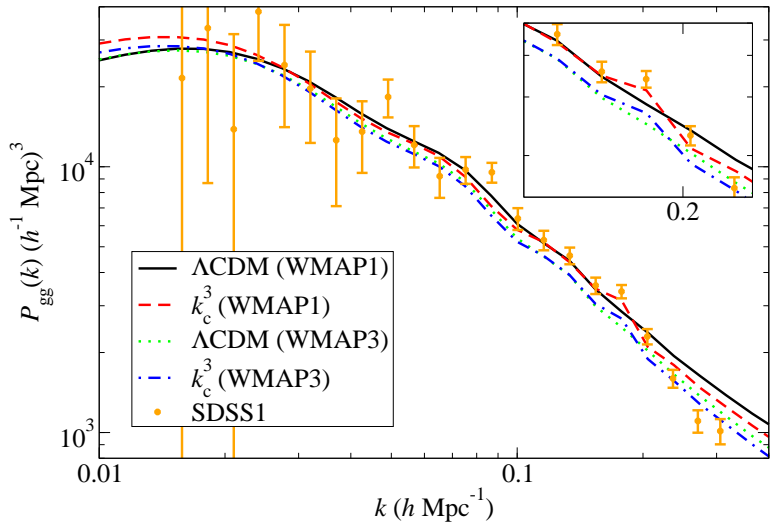}}
\caption{The best fit galaxy-galaxy power spectra of WMAP + SDSS1
for the power-law $\Lambda$CDM model and the $k^3_{\rm c}$ model
(going up to $k/h=0.2\ {\rm Mpc}^{-1}$). The SDSS1 data are also shown.}
\label{mpk_wsnon}
\end{figure}

\Fig{mpk_wsnon} shows the best fit galaxy-galaxy power spectra of
WMAP + SDSS1 for the power-law $\Lambda$CDM and $k^3_{\rm c}$ models using nonlinear theory
(fitting to $k/h=0.2\ {\rm Mpc}^{-1}$). The vertical shifts between
the results of WMAP1 and WMAP3 are probably due to the shift of $n_s$
from 0.959 to 0.943. The bumps at $k/h\sim0.18\ {\rm Mpc}^{-1}$ are
believed to be the results of the $k^3$ peaks in the same regime.
Therefore, our results show that, except for the low ($k \sim 0.004\ {\rm
Mpc}^{-1}$) and high ($k \sim 0.04\ {\rm Mpc}^{-1}$) multipoles, the $k^3$
component could also appear in the nonlinear regime ($k \sim 0.14\ {\rm
Mpc}^{-1}$). However, since it is not theoretically
well-understood how a power spectrum with a $k^3$ component goes
nonlinear, further investigation is needed before drawing any
firm conclusions.

Table \ref{tab1} also shows the results of fitting WMAP and SDSS4. 
The $k^3$ component can fit the data well ($\Delta{\chi^2}^\prime=2.4$ for WMAP3 + SDSS4). However, the bumps caused by the $k^3$ component are not as obvious as in the SDSS1 samples, so the power spectra of SDSS4 are not shown.
The total $\Delta\chi^2$'s are smaller in the case of SDSS4 because the
LRG sample favors a lower cutoff, $k \sim 0.08\ {\rm Mpc}^{-1}$, 
where the WMAP data are better fit by a nearly scale-invariant spectrum.

We emphasize that when the nonlinear SDSS data are included, one can
still adjust the $k^3$ component to appear in the WMAP regime. For example, the global best fit model for WMAP3 + SDSS1 has $\Delta\chi^2 = 2.1$ (though $\Delta{\chi^2}^\prime=-0.4$), which is slightly better than for the best fit model where the $k^3$ component appears in the nonlinear regime ($\Delta\chi^2=0.4$ and $\Delta{\chi^2}^\prime=2.4$).

\subsection{Fitting the Lyman-$\alpha$ Data}
\label{lyman-alpha}

The Lyman-$\alpha$ forest probes the highest-$k$ regime, and it usually
has a large effective redshift ($z>2$). Therefore, even though it is probing
smaller scales, the scales are still in the linear regime because it is
looking at early times (smaller scales go nonlinear first).

\begin{figure}[ht]
\centering{\includegraphics[width=\textwidth]{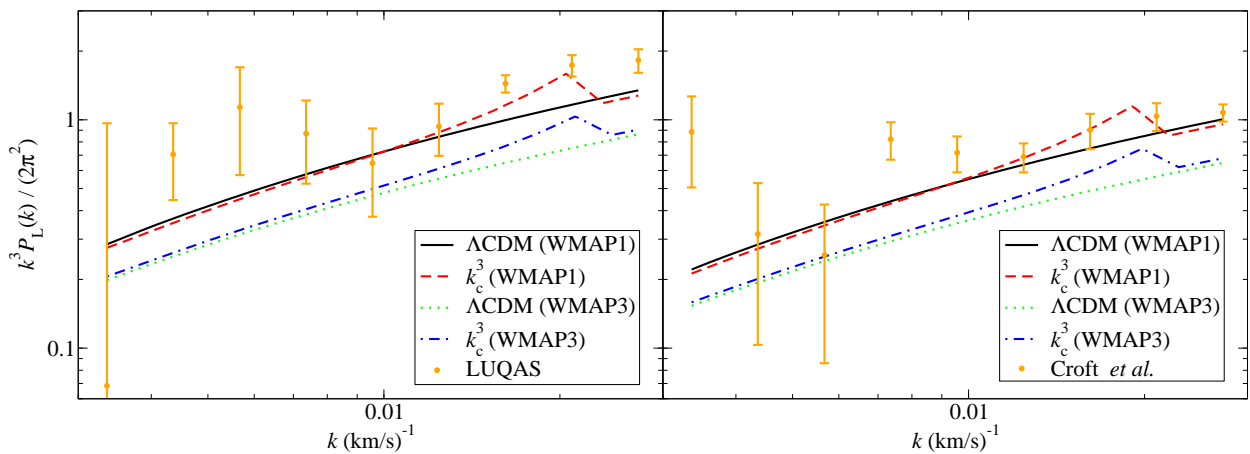}}
\caption{The best fit linear power spectra of WMAP + Viel {\it et al}.\ Lyman-$\alpha$ for the power-law $\Lambda$CDM model and the $k^3_{\rm c}$ model. The LUQAS ($z=2.125$) and Croft {\it et al}.\ ($z=2.72$) data are also shown.}
\label{lya_viel}
\end{figure}

As mentioned in Section \ref{data}, we use two different
Lyman-$\alpha$ data sets to test the $k^3$ component: Viel {\it et
al}.\ \cite{Viel, Kim, Croft} and SDSS Lyman-$\alpha$ \cite{sdsslya1,
sdsslya2} data. The lower part of Table \ref{tab1} lists the best fit
$k^3_{\rm c}$ models of WMAP + Viel {\it et al}.\ Lyman-$\alpha$.
\Fig{lya_viel} shows the corresponding best fit linear power spectrum
at $z=2.125$ and 2.72, respectively. The results of WMAP1 and WMAP3
are consistent except for the vertical shifts, which are probably due
to the change of the spectral index from 0.973 to 0.959.  In this
case, the improved fit ($\Delta {\chi^2}^\prime = 3.8$ for WMAP3 + Lyman-$\alpha$) is due to accounting for a bump at $\sim 0.02\
{\rm (km/s)}^{-1}$ in  the observed power, which corresponds to a 
$k^3$ peak in the primordial power spectrum near $\sim 1.5\ {\rm
Mpc}^{-1}$. The best fit $k^3_{\rm c}$ model for WMAP3 + Viel {\it et
al}.\ has matter density $\Omega_m = \Omega_b+\Omega_c=0.27$,  so the
correspondence between the two different ways of specifying the wave
number is given by  1 (km/s)$^{-1} \sim 97h\ {\rm Mpc}^{-1}$ for
$z=2.125$ and  1 (km/s)$^{-1} \sim 104h\ {\rm Mpc}^{-1}$ for 
$z=2.72$ \cite{LiddleLyth}. As in the case of WMAP3 + SDSS1
(nonlinear), the global best fit model of WMAP3 + Viel {\it et al}.\
has the $k^3$ peak appear in the WMAP regime, which gives
$\Delta\chi^2=5.7$ ($\Delta{\chi^2}^\prime=0.2$).

We also test the $k^3_{\rm c}$ model using the SDSS Lyman-$\alpha$
data. It turns out that there is no similar evidence for the $k^3$
component as in the case of WMAP + Viel {\it et al}., i.e. the maximum ratio $r_3(k_{\rm c})$ is of order $o(0.01)$,
so the $k^3_{\rm c}$  model is not distinguishable from the power-law
$\Lambda$CDM model and hence the $\Delta\chi^2$ is of order
$O(0.1)$.  At first glance it is a bit surprising that adding two new
parameters does not improve the fit substantially. However,  the SDSS
Lyman-$\alpha$ data are not given in a band-power description of
$P_{\rm L}(k)$; instead, the information is extracted in terms of the
amplitude, $k^3P_{\rm L}(k,z)/2\pi^2$, its tilt, $n_{\rm eff} = d\ln
P_{\rm L}(k)/d\ln k$, and the running, $\alpha_{\rm eff} = dn_{\rm
eff} / d\ln k$, at a pivot redshift $z_p=3.0$ and a pivot wavenumber
$k_p=0.009\ ({\rm km/s})^{-1}$ \cite{sdsslya1, sdsslya2}. It is not
obvious that such a description of the data can accommodate the extra
freedom available in the $k^3_{\rm c}$ model.

\subsection{Constraint on the Amplitude Ratio}
\label{constraint}

In the absence of significant positive evidence for a $k^3$ distortion, we now consider the experimental
limit on the magnitude of
the $k^3$ component in the primordial power spectrum.
The right panels of \fig{w3_k3_problike} show the
two-dimensional marginalized probabilities (and their 68\% and
95\% CL contours) and mean likelihood of samples of $\ln r_3$ and $\ln k_{\rm
c}$ for the $k^3_{\rm c}$ model using the WMAP3, SDSS1 (linear), and Viel {\it et al}.\ Lyman-$\alpha$
data. There are three peaks in the mean likelihood of samples, and they correspond the local best fits of WMAP (low and high multipoles) and Lyman-$\alpha$. But only the peak
of WMAP (high multipoles) in the marginalized probability can be seen, since it has a
large volume of samples.

Again, the waterfall-like surfaces indicate
that the high $\ln r_3$ and $\ln k_{\rm c}$ region is ruled out by
the data, but the one-dimensional marginalized constraints on $\ln
r_3$ and $\ln k_{\rm c}$ are meaningless, because of the
strong mutual dependence of the two parameters.
However, the constraint on $r_3(k_{\rm c})$ can be inferred 
from the two-dimensional marginalized contours. From \eq{P_Rk3}, 
the maximum amplitude ratio is
\be
	r_3(k_{\rm c}) = r_3(k_0)\left(\frac{k_{\rm c}}{k_0}\right)^{4-n_s},
\ee
so we have
\bea
	\ln r_3(k_{\rm c}) &=& \ln r_3(k_0) + (4-n_s) \ln k_{\rm c} - (4-n_s)\ln k_0\nonumber\\
	&\simeq& \ln r_3(k_0) + 3 \ln k_{\rm c} - 3\ln k_0,
\eea
where we have used the fact that the current best fit model has $n_s \simeq 1$. Therefore, plotting a series of lines with approximately the same $r_3(k_{\rm c})$,
\be
	\ln r_3(k_0) \propto - 3 \ln k_{\rm c},
\ee
and finding their intersection with the two-dimensional marginalized 
contour gives the best bound on the maximum amplitude ratio. 
From the lower part of \fig{w3_k3_problike} (the dashed, red lines), we find that
\be
	r_3 < 0.7\ (95\% \ {\rm CL})\hskip0.25in{\rm for}\hskip0.25in 1.8\times10^{-3}\ {\rm Mpc}^{-1} < k < 0.3\ {\rm Mpc}^{-1}
\ee
when using WMAP3 alone, and 
\be
	r_3 < 1.5\ (95\% \ {\rm CL})\hskip0.25in{\rm for}\hskip0.25in 2.3\times10^{-3}\ {\rm Mpc}^{-1} < k < 8.2\ {\rm Mpc}^{-1}
\ee
when using WMAP3 + SDSS4 (linear) + Viel {\it et al}.\ Lyman-$\alpha$.

It is possible that for a heavy tachyon, the cutoff on its 
perturbations lies below the smallest currently observable scales. Hence,
we also investigate the model without a cutoff, i.e.\ the $k^3$ model. From
the point of view of the algorithm, this is equivalent to putting the
cutoff below the smallest scales in the data or, equivalently, above the
largest $k$ value called by CosmoMC. We tested the $k^3$ model with all
the joint data used in the previous sections. All of the results give a
negligible amplitude ratio, i.e.\ $\ln r_3 \to -25$
(the lower prior), making the model indistinguishable from the power-law
$\Lambda$CDM model.
This gives rise to the following observational constraint on the $k^3$ model. The upper limit of the magnitude of the $k^3$ term can be read directly from the bottom-right panel of \fig{w3_k3_problike}; taking the largest cutoff value (the upper prior is $k=15\ {\rm Mpc}^{-1}$ when Lyman-$\alpha$ is used) in the contour gives
\be
	r_3 (15\ {\rm Mpc}^{-1}) < 11\ (95\%\ {\rm CL}).
\ee

\section{Theoretical Models}
\label{models}
\subsection{Hybrid Inflation}
In the previous sections we have shown that bumplike features in 
the power spectra of the CMB and LSS can be
explained by the addition of a $k^3$ contamination of the nearly
scale-invariant contribution.  We noted in the introduction that
preheating at the end of chaotic inflation predicts $k^3$ components
which are too small to be observed on the CMB-LSS scales
\cite{LLMW, FK, STBK}, but recent work has found that observable
effects can be generated for some ranges of parameters in hybrid
inflation theories  \cite{k3, k3_2}, which are defined by the
potential \cite{Linde}
\be
\label{pot}
     V(\varphi,\sigma) = 
   \frac{\lambda}{4} \left( \sigma^2 - v^2 \right)^2 
   + \frac{m_\varphi^2}{2}\varphi^2 + \frac{g^2}{2} \varphi^2 \sigma^2,
\ee
where $\varphi$ is the inflaton and $\sigma$ is the tachyonic field.

Inflation starts with large values of $\varphi$ and 
ends after the field-dependent tachyon mass parameter
becomes negative:
\be
\label{msigma}
	m^2_\sigma = -\lambda v^2 + g^2\varphi^2 < 0.
\ee
For certain ranges of the parameters $v$, $\lambda$, and $g$, the fluctuations
in $\sigma$ can become so large that their contribution to the
curvature perturbation, at second order in cosmological perturbation
theory, exceeds the usual first-order contribution due to the
inflaton $\varphi$.  The growth of fluctuations gets cut off at some
number $N_*$ of e-foldings after the onset of the tachyonic
instability, when their energy density starts to exceed the false
vacuum energy that drives inflation.  At this point, inflation ends
and there is no further evolution in the curvature perturbation on 
superhorizon scales.  Even if $N_*$ represents a very
short amount of time during which the fluctuations can grow in
magnitude, they grow exponentially fast, and so it is possible to 
achieve a large effect.  

Without going into all the details, one can still appreciate
why it is possible to obtain a spectrum of the form $k^3$ with
a cutoff. In fact, the $k^3$ spectrum is completely generic for
massive fields, i.e.\ with $m \gg H$, where $H$ is the
Hubble scale during inflation.  This can be found by solving
the Klein-Gordon equation for linearized fluctuations of the
fields in de Sitter space; it is straightforward to show that
$\langle (\delta\sigma)^2\rangle \sim m^{-1} \int d^{\,3} k$
for heavy fields, whereas  $\langle (\delta\varphi)^2 \rangle
\sim H^2 \int d^{\,3} k/k^3$ for light ($m\ll H$) fields. However,
because of the redshifting of the physical wave number, a $k^3$
component in the spectrum is usually negligible compared to the 
scale-invariant contribution.  Only if the $k^3$ component is
amplified can it become observable.  Thus, the
other relevant effect is the amplification of the tachyonic
fluctuations by the instability, which causes modes with $k^2 < a^2
|m^2_\sigma|$ to grow as $\exp [t\sqrt{ |m^2_\sigma|-(k/a)^2}]$.  We
see that there is naturally a cutoff  $k_{\rm c} = a |m_\sigma|$ on the
modes which get amplified.  Of course, this scale gets
stretched by the Hubble expansion which occurs after
inflation.  At the present time, the cutoff scale is therefore
given by 
\be 
\label{kc}
   k_{\rm c} = |m_\sigma| {H_0\over H_i} {\rm e}^{N_{\rm e}} ,
\ee where
$H_0$  and $H_i$ are respectively the Hubble rate today and
during inflation, and $N_{\rm e}$ is the number of e-foldings of
inflation since the horizon crossing, $N_{\rm e}\sim 60$.  (In the
present analysis, we do not assume that $N_{\rm e}\sim 60$; rather, we
determine $N_{\rm e}$ according to the actual reheat temperature,
assumed to be of the order $\lambda^{1/4} v$ since $\lambda
v^4$ is the false vacuum energy density which drives
inflation.)  Notice that if $|m_\sigma|=H_i$, $k_{\rm c}/H_0$ is just
the ratio of scales which crossed the horizon at the beginning
and at the end of inflation, as expected.  In order to match
the bumps in the WMAP data, we need $k_{\rm c}/H_0$ to be on the order
of 16 or 162.  Due to the large factor ${\rm e}^{N_{\rm e}}$ in \eq{kc}, this is difficult to achieve.  However, one should
recall that $m_\sigma$ is field-dependent (\eq{msigma}),
and it passes through zero shortly before the end of inflation.
For some values of parameters, $m_\sigma$ can still be small
at the end of inflation; moreover, $N_{\rm e}$ can also be relatively
small if the reheat temperature is low.  It is therefore not
obvious whether small enough values of $k_{\rm c}$ can be achieved
to match the phenomenological requirements that were suggested
by the previous sections.

\begin{figure}[htp]
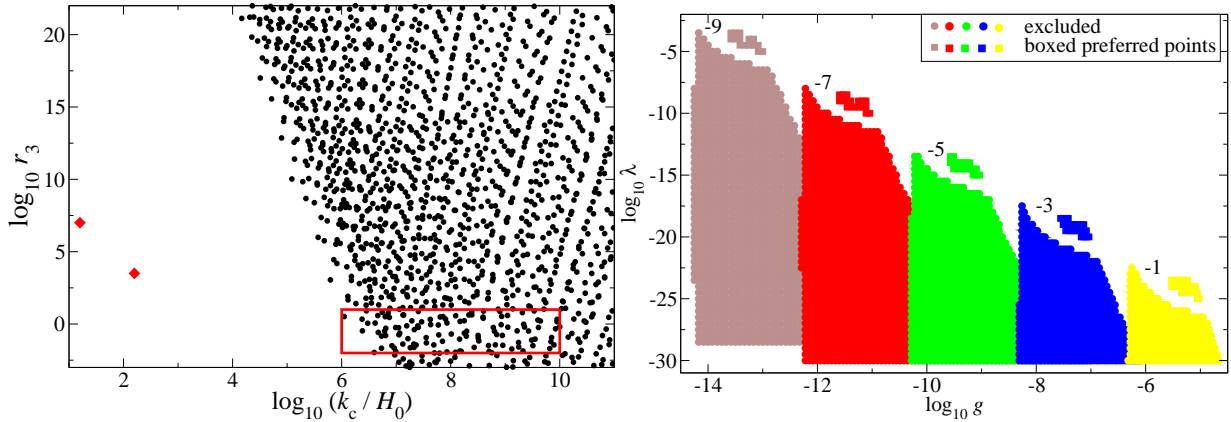

\centering{\includegraphics[width=3.2in]{k3-rc_4.eps}\includegraphics[width=3.2in]{k3-par_3.eps}}
\caption{Left: values of $r_3$ (at $k = 1$ Mpc$^{-1}$)
 and $k_{\rm c}$ which are achieved
by varying parameters of the hybrid inflation model.
The diamonds are the best fit models of WMAP at low and high multipoles; 
the boxed area is the Lyman-$\alpha$ preferred 
region $-2< \log_{10}r_3< 1$ and
$6< \log_{10}(k_{\rm c}/H_0) < 10$. Right: preferred points (square) next to excluded regions
(circles) for fitting Lyman-$\alpha$ data in the hybrid inflation
model. The numbers ($-1,-3,-5,-7,-9$) indicate the value of
$\log_{10} v$ for the respective regions.}
\label{k3h}
\end{figure}

To investigate whether the parameters suggested by our fits can
correspond to hybrid inflation, we have adapted the code used
in Refs. \cite{k3, k3_2}, which numerically computed
the magnitude $r_3$ of the $k^3$ contribution, scanning
over a large range of the model's parameters, $g$, $\lambda$,
and $v$.  In the left panel of \fig{k3h} we plot the values of
$r_3$  (at $k = 1$ Mpc$^{-1}$) versus $k_{\rm c}$ obtained by scanning
over the parameter values $\log_{10} v = -1,-3,-5,-7,-9$,
$\log_{10}g^2 > -30$, and $\log_{10}\lambda < 0$. We find no
cases where the peak parameters $k_{\rm c}$ and $r_3$ are in the
range needed for the WMAP glitches at $\ell \sim 10 - 50$ and 
$\ell \sim 540$ (the diamonds in the figure). The left panel of the figure shows that points having small
enough values of $k_{\rm c}$ all have overly large values  of
$r_3$.  However, there are realizations in which $k_{\rm c}$
is close to the 1 Mpc$^{-1}$ suggested by the Lyman-$\alpha$
data, as well as points where  $k_{\rm c}$ is larger and is 
effectively removed from the analysis (the $k^3$ model).
Such points are contained in the narrow box shown in the
lower right-hand corner of the left panel of \fig{k3h}.

The promising points in the left panel of \fig{k3h} (those falling within the box)
correspond to values of the hybrid inflation parameters shown
in the right panel of the figure.  These are shown as boxes in the $g$-$\lambda$
plane, for the values of $\log_{10} v = -1,-3,-5,-7,-9$,
along with excluded regions based on the previous analysis
of Refs.\ \cite{k3, k3_2}, which demanded that the $k^3$ effect 
not be too large.  As expected, the points suggested by our
current fits to the data are close to the boundaries of the 
excluded regions.

\subsection{Double D-Term Inflation}
\label{DDT}

We have seen that pure hybrid inflation is able to give a sizeable
$k^3$ component, but with a cutoff $k_{\rm c}$ larger than the ranges
observable in LSS and Lyman-$\alpha$ data.  A more complicated 
but still well-motivated class of models is able to bring $k_{\rm c}$
into the observable part of the spectrum, as was shown by Ref.\
\cite{julien1}.\footnote{We thank Julien Lesgourgues
for calling our attention to this work.}  These are double inflation models of the D-term
hybrid type \cite{sak, julien2, kanazawa}, based on the superpotential
\be
	W = \alpha\ A A_+ A_- + \beta B B_+ B_-,
\ee
with six superfields $A$, $A_\pm$, $B$, $B_\pm$, with charges
$(g_A,g_B)$ under two $U(1)$ gauge groups $U(1)_A$ and $U(1)_B$,
given by $(0,0)$ for $A,B$, $(\pm 1,0)$ for $A_\pm$, and $(\pm1,\pm1)$
for $B_\pm$.  It leads to a scalar potential of the form
\be
	V = {g_A^2\over 2}\left(\xi_A-\frac12 |C|^2\right)^2
	+\frac14\beta B^2 |C|^2+
	{g_B^2\over 2}\left(\xi_B-\frac12 |C|^2\right)^2,
\label{Vdterm}
\ee
plus one-loop corrections (depending on $A,B,C$), where $\xi_i$ are
Fayet-Iliopoulos terms, $A$ and $B$ are proportional to the modulus
of the scalar component of the corresponding superfields, and $C$ is
the complex scalar component of $B_-$.  

\begin{figure}[htp]
\centering{\includegraphics[width=4in]{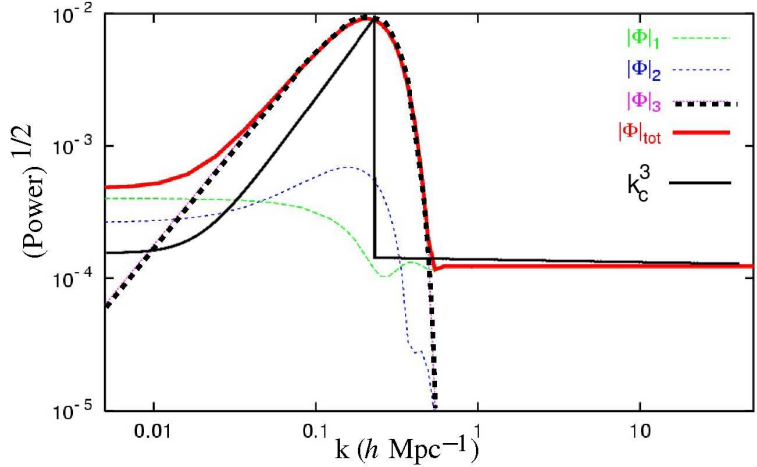}}
\caption{A sample power spectrum from double D-term inflation, adapted
from Ref.\ \cite{julien1}. The bold dashed line is the $k^3$ contribution
from tachyonic fluctuations. The model parameters were $\sqrt{\xi_A} =
3\times10^{-3}M_{\rm Pl}$, $\sqrt{\xi_B} = 4.2\times10^{-3}M_{\rm Pl}$,
$g_A=g_B=10^{-2}$, and $\beta=0.5\times10^{-3}$. For comparison,
a $k^3_{\rm c}$ model 
is also shown, with parameters
$r_3(k_0)=1.3\times10^{-3}$ and $k_{\rm c}=0.3h\ {\rm Mpc}^{-1}$.}
\label{julien}
\end{figure}

From the form of (\ref{Vdterm}) it can be seen that the field $C$
becomes tachyonic when $B$ rolls down to a critical value.  However,
in this model, the instability of $C$ need not trigger the end of
inflation; instead, there can be a second stage via rolling of the
field $A$.  The fluctuations due to the tachyonic preheating are
found to have a  $k^3$ spectrum, and a sharp cutoff at the scale
which crossed the horizon during the phase transition.  A $k^3$ spike
is generated, as shown in \fig{julien}.  In this example, the
spike is located at rather high $k$ values, but there is plenty of
freedom in the model to move this to arbitrarily lower scales.  One
merely needs to adjust the duration of the second stage of inflation 
so that the phase transition occurs sufficiently soon after horizon
crossing of modes which are at the present horizon.  This model is
therefore a good candidate for generating the bumps in the low
multipoles which we have investigated in Section \ref{fittingwmap}.
We also show a $k^3_{\rm c}$ power spectrum in \fig{julien}. 
By appropriate choices of the parameters
of the double D-term inflation, the  wavenumber and the amplitude of
the feature can be adjusted.  However, the most closely corresponding
spike in the $k^3_{\rm c}$  model has a narrower width, so one would
have to add another parameter to the $k^3_{\rm c}$ ansatz so as to
accurately mimic the prediction of the double D-term inflation scenario.

\section{$k^n$ Features  in the Primordial Power Spectrum}
\label{knspike}

As mentioned in the previous sections, the $k^3_{\rm c}$ model is theoretically well motivated, and we take the point of view that it is a good representative of the generic spikelike feature in the primordial power spectrum. To clarify this point, we compare the $k^3_{\rm c}$ model to the $k^n_{\rm c}$ models with nearby values ($n=1,2,4,5$). Since the evidence from LSS and Lyman-$\alpha$ data is moderate, we focus on the WMAP3 data. Table \ref{tab3} shows the best fit models for different $k^n$ spike power spectra.

\begin{table}[htp]
\caption{The best fit $k^n_{\rm c}$ models of WMAP3, fitting to the low multipoles ($\ell \sim 10 - 50$, upper) or the second peak ($\ell \sim 540$, lower) of the CMB angular spectrum.}
\label{tab3}
\renewcommand{\arraystretch}{1.5}
\vspace{2ex}
\centering{\begin{tabular}{ccccc}
\hline\hline
Model & $\Delta\chi^2$ & $n_s(k_0)$ & $k_{\rm c}$ ($10^{-2}\ {\rm Mpc}^{-1}$) & $r_n(k_{\rm c})$\\
\hline
$k^1$ & 0.2 & 0.970 & 0.434 & 0.159\\
$k^2$ & 1.2 & 0.985 & 0.384 & 0.446\\
$k^3$ & 2.0 & 0.969 & 0.376 & 0.397\\
$k^4$ & 1.5 & 0.974 & 0.382 & 0.271\\
$k^5$ & 1.9 & 0.981 & 0.369 & 0.797\\
\hline
$k^1$ & 5.8 & 0.944 & 3.94 & 0.123\\
$k^2$ & 4.3 & 0.942 & 3.82 & 0.153\\
$k^3$ & 5.4 & 0.944 & 3.88 & 0.139\\
$k^4$ & 4.8 & 0.946 & 3.80 & 0.200\\
$k^5$ & 5.4 & 0.959 & 3.79 & 0.191\\
\hline\hline
\end{tabular}}
\end{table}

It can be seen from the table that all $k^n_{\rm c}$ models are consistent: a $k^n$ spike at $k\sim0.004\ {\rm Mpc}^{-1}$ can fit the low-multipole glitches ($\ell \sim 10 - 50$) and a $k^n$ spike at $k\sim0.04\ {\rm Mpc}^{-1}$ can improve the fits of the second peak ($\ell \sim 540$) of the CMB angular spectrum. While all $k^n$ spikes have roughly the same order of improvement ($\Delta\chi^2$), it is interesting that the $k^3_{\rm c}$ model is slightly favored by the data.

We emphasize that the power $n$ is fixed at $n=3$ in most of the investigation; only in this section so we allow it to change to some nearby values, to show that $n=3$ is a good representative of the generic spikelike feature.

\section{Conclusions}
\label{conclusions}

We have examined a class of perturbations  to the nearly
scale-invariant spectrum, coming from preheating, which is
motivated by very general principles of causality.   It gives a
$k^3$ component which adds to the usual $k^{n_s-1}$ spectrum,
up to some cutoff $k_{\rm c}$ which is determined by the microphysics
of preheating.  We did a phenomenological analysis of this kind
of distortion to the primordial power spectrum, using the 
Monte-Carlo Markov-chain algorithm provided by CosmoMC, and
found that such a component could  improve either the fits to
the irregularities of the low multipoles ($\ell \sim 10 - 50$) or
the second peak ($\ell \sim 540$) of the CMB spectrum,  giving 
$\Delta\chi^2=3.6$ or 1.2 at the low multipoles and
$\Delta\chi^2=2.0$ or 5.4 at the second peak, using the  WMAP1
or WMAP3 data, respectively. Moreover, the amplitude of the
$k^3$ contribution was found to  be as large as 0.69 or 0.14 of
the $k^{n_s-1}$ part when fitting the low multipoles or the
second peak, respectively. These results provide an intriguing
suggestion for such a $k^3$ component, but more generally they
indicate that the WMAP data are consistent with sizable
deviations from a nearly scale-invariant spectrum.

We also studied the CBI and ACBAR data to investigate whether
the $k^3$ component could explain excess power in the high
multipoles seen by those experiments. Our results show that the
CBI and ACBAR data are consistent with a $k^3$ component at the
low multipoles or the second peak, but they do not give
evidence for a $k^3$ component at the high multipoles; in
retrospect this had to be the case, due to Silk damping.
However, this limitation does not apply to LSS
data, i.e.\ 2dFGRS and SDSS, which explore a similar
range of $k$ space as CBI and ACBAR. Again, the LSS data
are consistent with the results from WMAP, but there is no
evidence for the $k^3_{\rm c}$ model at $k/h < 0.15\ {\rm
Mpc}^{-1}$ ($0.1\ {\rm Mpc}^{-1}$ for SDSS4).

We also tested the $k^3_{\rm c}$ model in the nonlinear regime
of the galaxy-galaxy spectrum, $k/h > 0.15\ {\rm Mpc}^{-1}$ ($0.1\ {\rm Mpc}^{-1}$ for SDSS4). 
We found that the $k^3$ component can tune the shape of the
SDSS galaxy-galaxy power spectrum, giving an improvement of
$\Delta\chi^2 = 0.9$ ($\Delta{\chi^2}^\prime = 2.2$) for SDSS1; SDSS4 favors a lower cutoff, $k \sim 0.8\ {\rm Mpc}^{-1}$, which gives $\Delta\chi^2 = 0.4$ ($\Delta{\chi^2}^\prime=2.4$).
However, we feel that the implementation of
nonlinear evolution in the likelihood code for SDSS has not yet
been tested thoroughly in conjunction with a nonstandard
spectrum such as the one we are using, so we reserve judgment
as to these particular results.  They should
be taken as motivation for a more detailed study of the
nonlinear regime.

A further handle on the power spectrum at high $k$ is provided
by the Lyman-$\alpha$ forest data.
We found that the Viel {\it et al}.\ data allow a large
amplitude ratio (0.33), with an
improvement of $\chi^2=2.5$ ($\Delta{\chi^2}^\prime=3.8$).
The SDSS Lyman-$\alpha$ data, however, do not give similar evidence in the same regime.

In brief, the CMB, LSS, and Lyman-$\alpha$ data are
consistent with a nearly scale-invariant spectral index plus a
$k^3$ component. 
We further showed that the $k^3$ spike is a good representative
of the generic spikelike feature in the primordial power spectrum
by comparing different $k^n_{\rm c}$ models.
By adjusting both the location and the amplitude
of the $k^3$ component, one can of course always find better
fits.
We emphasize that due to the addition of two free
parameters, the evidence for the $k^3$ component is not compelling.
However, our results show that the $k^3_{\rm c}$ contamination is
not ruled out by the data, even when its amplitude is surprisingly
large. We have determined  constraints on the magnitude of this extra
component, finding an  upper limit of  $r_3<1.5$ (95\% CL) on the
amplitude ratio, over the range of wave numbers $2.3\times10^{-3}\
{\rm Mpc}^{-1} < k < 8.2\ {\rm Mpc}^{-1}$.

We have also explored in some detail the parameter space of the 
hybrid inflation model, which was found to give an observably large
$k^3$ component for some ranges of the model's parameters.  The model
is able to match the observations in the cases where the benefit of
the $k^3$ component is less rigorously shown, namely in the
highest-$k$ regions of the spectrum. For the features at lower $k$,
the double D-term inflation model discussed in Section \ref{DDT}
appears to be ideally suited for generating spikes at the large
scales we have investigated here.

\section*{Acknowledgements}

We thank Gilbert Holder for his collaboration in the initial stages of
this work, Neil Barnaby for helpful discussions, and Antony Lewis,
An\v{z}e Slosar, and Patrick McDonald for help with CosmoMC and its
SDSS Lyman-$\alpha$ patch. We are grateful to Julien Lesgourgues for
constructive criticism of the manuscript. Loison Hoi is supported by
the Dow-Hickson Fellowship in Physics at McGill University. We are
also supported by NSERC of Canada and FQRNT of Qu\'ebec.

\appendix
\section{Relation of the $k^3$ Spectrum to the Causality Constraint}
\label{appendixa}
Due to changes of notation over the years, readers of the 
original causality argument due to 
Abbott and Traschen \cite{Abbott-Traschen} may not immediately
perceive that their result implies the $k^3$ spectrum for causal
perturbations.  In this appendix we clarify the relation.
Reference \cite{Abbott-Traschen} assumes that
$\delta\rho(\mf x)/\rho = \sum_a c_a F_a(\mf x - \mf x_a)$, where
for convenience $c_a$ is a random variable which obeys
$\left\langle c_a c_b\right\rangle = c^2 \delta_{ab}$.  The correlation function
in position space is
\be
	\left\langle{\delta\rho(\mf x)\over\rho} 
	{\delta\rho(\mf y)\over\rho}  \right\rangle = c^2\sum_a
	F_a(\mf x - \mf x_a) F_a(\mf y - \mf x_a).
\ee
In Fourier space,
\be
	\left\langle{{\delta\rho}_{\mf k}^*\over\rho} 
	{{\delta\rho}_{\mf k^\prime}\over\rho}  \right\rangle = c^2\sum_a
	\mbox e^{i(\mf k - \mf k^\prime)\cdot \mf x_a} k^2 {k^\prime}^2,
\ee
where we assumed that each $F_a$ has Fourier transform $F_{a\mf k} \sim |\mf k|^2 = k^2$, which
is the essential restriction due to causality.  If there
are enough random centers $\mf x_a$, then the sum will approximately give
a delta function,
\be
	\left\langle{{\delta\rho}_{\mf k}^*\over\rho }{{\delta\rho}_{\mf k^\prime}\over\rho }\right\rangle \simeq c^2 \left({2\pi \over L}\right)^3k^2{k^\prime}^2\delta_{\mf k\mf k^\prime},
\ee
where $L$ is a box size.
Therefore, one has
\be
	\label{res1}
	\left\langle \left| {{\delta\rho}_\mf k \over \rho}\right|^2\right\rangle \simeq c^2 \left({2\pi \over L}\right)^3k^4.
\ee
The $k^4$ behavior is potentially confusing, but we will now
show that this translates into $k^3$ behavior for the 
power spectrum of the curvature
perturbation  ${\cal R}$.  

The relation between
${\cal R}_\mf k$ and ${\delta\rho}_\mf k/\rho$ is \cite{LiddleLyth}
\be
\label{res4}
{{\delta\rho}_{\mf k}\over\rho} = \frac25\left(k\over aH\right)^2{D_1(\Omega_m) \over \Omega_m}T(k,\Omega_{m0}){\cal R}_\mf k,
\ee
where $D_1(\Omega_m)/\Omega_m$ is the growth factor for the matter perturbation and $T(k,\Omega_{m0})$ is the transfer function. Since the growth factor does not have $k$ dependence and the causality argument applies to the limit $k\to0$ where
$T(k)$ is roughly a constant,
\be
\label{res2}
\left\langle\left|{{\delta\rho}_{\mf k}\over\rho }\right|^2\right\rangle \sim k^4
\left\langle\left|{\cal R}_{\mf k}\right|^2 \right\rangle.
\ee
On the other hand, the power spectrum is defined through
\be
\label{res3}
	{\mathcal {P_R}}(k) \equiv \left({L \over 2\pi}\right)^3 4\pi k^3 \left\langle\left|{\mathcal R}_\mf k\right|^2\right\rangle.
\ee
\Eqs{res2}{res3} imply that
\be
	\left\langle\left|{{\delta\rho}_\mf k\over\rho }\right|^2\right\rangle \sim
	\left({2\pi \over L}\right)^3{k\over 4\pi}\mathcal{P_R}(k).
\ee
Making a comparison with \eq{res1}, we see that
\be
	{\cal P_R}(k) \sim k^3
\ee
if $\delta\rho$ has the assumed behavior.

The typical onset value for $T(k)$ to decrease is of order $0.01\ {\rm Mpc}^{-1}$. Therefore, if the causality constraint is going to give a $k^3$ spectrum, then the $k^3$ component appears only at $k<0.01\ {\rm Mpc}^{-1}$, otherwise the causality constraint will give a power greater than 3. Of course, the decrease of the $k^3$ component could be more complicated than a sharp cutoff.


\end{document}